\documentclass[epj]{svjour}

\usepackage{graphicx}
\usepackage{color}
\usepackage{amsmath,amssymb}
\usepackage{hyperref}

\definecolor{brown}{rgb}{0.6,0.3,0}
\definecolor{purple}{rgb}{0.3,0.0,0.6}

\newcommand\vb{V_{\rm sd}}
\newcommand\vg{V_{\rm g}}

\begin{document}

\title{Hund and pair-hopping signature in transport properties \\ of degenerate nanoscale devices}

\author{J. Azema, A.-M. Dar\'e, P. Lombardo}\email{pierre.lombardo@univ-amu.fr}
\authorrunning{J. Azema {\it et al.}}
\institute{Aix-Marseille Universit\'e CNRS, IM2NP UMR 7334, 13397 Marseille, France\\
\email{pierre.lombardo@univ-amu.fr}}

\date{\today}

\abstract{
We investigate the signature of a complete Coulomb interaction in transport properties of double-orbital nanoscale devices.  
We analyze the specific effects of Hund exchange and pair hopping terms, calculating in particular stability diagrams.
It turns out that a crude model, with partial Coulomb interaction, may lead to a misinterpretation of experiments.
In addition, it is shown that spectral weight transfers induced by gate and bias voltages strongly influence charge current. 
The low temperature regime is also investigated, displaying inelastic cotunneling associated with the exchange term, as well as  Kondo conductance enhancement.
}
\maketitle

\section{Introduction}
The strong repulsive Coulomb interaction is responsible for most of the nanoscale device properties like Coulomb blockade, or Kondo effect at low temperature~\cite{HKP07,NCL2000,P2002}. Local impurity Anderson models~\cite{A61,MWL1993} are then good candidates to investigate transport properties of these highly confined and strongly correlated systems.
Details of  local spectroscopic properties are important, particularly since transport channels may consist of sophisticated structures~\cite{TAH96}.
Consideration of this complexity is essential for a proper interpretation of the transport properties.
It allows for example a manipulation of the transition-metal-complex spin states~\cite{OMZ2010}. 
Also in single-wall carbon nanotube quantum dots, this complexity is responsible for Hund's coupling which directly affects the excitation spectroscopy measurements~\cite{MFS05}.
Local impurity Anderson model has therefore to be generalized \cite{KSO08,LWG09,FFR11,SNO12} to multi-orbital systems. 
%

%{\color{blue} 
However, in this context of orbitally degenerate Anderson model for magnetic impurities or quantum dots, different explicit choices for the Hamiltonian have led to different predictions for the Kondo temperature as a function the Hund's coupling~\cite{NH12}. 
Besides, in the context of electronic properties of compounds with transition metal ions, thorough discussions about the choice of the local Hamiltonian describing the correlated ions 
have been published~\cite{K1959,DHM01,OKH05}. 
Even in the simplest case $n=2$ of the $n$-orbitally degenerate model, the general Coulomb interaction contains many terms~\cite{hamiltonian}: in addition to intra $U$ and inter orbital $U'$ repulsions, it displays Hund's exchange term $J$ as well as a {\em pair hopping} parameter $J^\prime$~\cite{pairhopping}. 
This last term is far from being prevalent in literature: to simplify the calculations, $U'=U$ and $J'=0$ are often assumed, with few exceptions as in Ref.~\cite{WM2010,YWD2011}.  
Nevertheless, a recent theoretical work concerning lattice multiorbital Mott systems~\cite{LP11} has shown that both $J$ and $J^\prime$ should be considered  to get the correct  spectral weight transfer (SWT) upon doping between the various structures composing the electronic density of states.
Moreover, the different parameters are not independent, for example, real orbitals lead to $J =J'$. In the case of $d-$type orbitals one has $U' = U - 2 J$. 
The local Coulomb Hamiltonian is invariant under arbitrary rotation in the spin space, however it is not the case for rotation in the orbital space. To enforce partially this invariance, which is legitimate for degenerate orbitals, one can choose $U'=U-J-J'$~\cite{DHM01}. For $J'=0$ this leads to $U' =U-J$, and provides  invariance under arbitrary rotation in orbital space.

In this paper, we emphasize the importance of taking into account the whole Coulomb Hamiltonian for out of equilibrium properties across a nanoscale device in the Coulomb blockade regime. 
Specifically, we show that the Coulomb terms $J$ {\em and} $J^\prime$ have a major impact on spectral density high energy structures which are involved in the finite bias transport properties.
SWT between these structures occurs not only upon varying the gate voltage, but also by changing the bias voltage. This transfer affects deeply the appearance of stability diagrams. 
At low temperature, inelastic cotunneling processes with a characteristic threshold are driven by the exchange term $J$.

We therefore provide characteristic signatures which can be  experimentally relevant for interpreting stability diagrams and excitation spectroscopy measurements.
To this purpose, we calculate local spectral density for a finite applied bias voltage $\vb$ through a doubly-degenerate quantum dot.  The calculation is performed in a fully nonlinear manner within the framework of the Keldysh Green's function formalism.  More precisely, we use an out-of-equili\-brium generalization of the non-crossing approximation (NCA) \cite{WM94,HKH98}, which enables the computation of transport properties in presence of finite voltage across the quantum dot. 

\section{Model and method}
The system under study consists of a doubly degenerate dot coupled to two uncorrelated source and drain contacts.  The corresponding Anderson~\cite{A61} model reads:
$$H=H_{\rm d}+H_{\rm r}+H_{\rm c}\, ,$$ 
where $H_{\rm d}$  is the two-orbital ($m=a,b$) local Hamiltonian: 
\begin{eqnarray}
%\label{Hamiltonian} 
&H_{\rm d}&=
\varepsilon_{0} \sum_{m,\sigma} c_{m \sigma }^{\dagger } c_{m \sigma } 
\,+\, U\sum_{m}n_{m\uparrow}n_{m\downarrow}-2J \hat{\bf{S}}_{a}\hat{\bf{S}}_{b} \nonumber \\
 &+&  (U^{\prime}-\frac{J}{2})\sum_{\sigma,\sigma^{\prime}}n_{a\sigma}n_{b\sigma^{\prime}} 
 +  J^{\prime}\sum_{m\neq m^\prime }c_{m\uparrow}^{\dagger}c_{m\downarrow}^{\dagger}c_{m^\prime\downarrow}c_{m^\prime\uparrow}
\nonumber 
\end{eqnarray}
$\varepsilon_{0}$ is the energy of the degenerate level tuned by the gate voltage $\vg=\varepsilon_{0}$,
the operator $\hat{\bf S}_{m}= (\hat S^{x}_{m},\hat S^{y}_{m},\hat S^{z}_{m})$ 
is the $1/2$-spin operator for orbital $m$. 
The parameter $U$ denotes the repulsion between electrons occupying the same orbital, $U^{\prime}=U-J-J^\prime$ is the repulsion between electrons
occupying different orbitals, $J$ is the Hund exchange term, and $J^{\prime}$ is the double hopping term~\cite{Jt}. This last term is systematically present when exchange interactions take place. In the following, calculations are made for  two cases: $(J'=J, U'=U-2J)$ and $(J'=0,U'=U-J)$, with $J = U/10$.
Note that the Hamiltonian $H_{\rm d}$ is also relevant for double-quantum dots, however there is no more reason to enforce  invariance  under rotation in the orbital space in this case, and $U^{\prime}=U-J-J^\prime$ is no longer required. 

The second term $H_{\rm r}$ describes the left (L) and right (R) leads in standard notations:
\begin{eqnarray}
H_{\rm r}&=& \sum_{{\alpha\in\{{\rm L,R}\}}\atop {m\in \{ { a,b} \},k,\sigma}}\varepsilon_{\alpha k}\,a_{\alpha k m\sigma }^{\dagger} a_{\alpha k m\sigma}
\,\mbox{.}
\nonumber 
\end{eqnarray}
Finally,  $H_{\rm c}$ accounts for the hybridization tunneling between dot and leads which -- we assume -- conserves orbital and spin quantum numbers~\cite{note}:
\begin{eqnarray}
H_{\rm c}&=& \sum_{{\alpha\in\{{\rm L,R}\}}\atop {m\in \{ { a,b} \},k,\sigma}} 
\left( t_{\alpha} c_{m\sigma }^{\dagger} a_{\alpha k m\sigma}+t_{\alpha}^*a_{\alpha k m\sigma }^{\dagger } c_{m \sigma }\right)
\,\mbox{.}\nonumber 
\end{eqnarray}

In our calculations, the total electrode density of states  $N_{\alpha}(\varepsilon)=N_{\alpha\uparrow}(\varepsilon)+N_{\alpha\downarrow}(\varepsilon)$  are Gaussian (half-width $D$) but the results depend very slightly of their particular shape, provided that their width is large compared to the other energy scales of the system. The total hybridization amplitude $\Gamma=\Gamma_{\rm L}(\overline{\mu})+\Gamma_{\rm R}(\overline{\mu})$ will be used as energy unit, where $\Gamma_{\alpha}(\varepsilon)=\pi t_{\alpha}^2 N_{\alpha}(\varepsilon)$ and  $\overline{\mu}=\frac{1}{2}(\mu_{\rm L}+\mu_{\rm R})$ is the average chemical potential.

In the out of equilibrium regime, a finite voltage bias $\vb=(\mu_{\rm L}-\mu_{\rm R})/e$ is applied symmetrically across the device and the corresponding nonlinear electrical current is given by
\begin{eqnarray}
I_{\rm d}&=&\frac{2e}{h}\int \left[f_{\rm L}(\varepsilon)-f_{\rm R}(\varepsilon)\right]\tau(\varepsilon)\,{\rm d}\varepsilon\, ,
\nonumber
\label{IQeqs}
\end{eqnarray}
where $h$ denotes Planck's constant and $-e$ the electronic charge. $f_{\alpha}(\varepsilon)\equiv f(\varepsilon-\mu_{\alpha})$ are the lead Fermi functions, and $\tau(\varepsilon)=\frac{\pi}{4} A(\varepsilon)\Gamma(\varepsilon)$, with $A(\varepsilon)=-\frac{1}{\pi}{\rm Im}[\sum_{m,\sigma}G_{m\sigma}(\varepsilon+i\delta)]$, the total dot spectral density. The retarded Green's function $G_{m\sigma}(\varepsilon+i\delta)=\langle\langle c_{m\sigma}; c_{m\sigma}^{\dagger}\rangle\rangle$ is obtained from a generalized Keldysh-based out-of-equilibrium NCA~\cite{WM94,HKH98}.
In the Coulomb blockade regime, the reliability of this approximation is guaranteed since the NCA is known to give accurate results for the spectral density down to temperatures of the order of a fraction of the Kondo temperature $T_{\rm K}$~\cite{NCArecent}.

\section{Results}
\subsection{Spectral densities  at equilibrium}
\begin{figure}
\begin{center}
    \includegraphics[width=8.5cm]{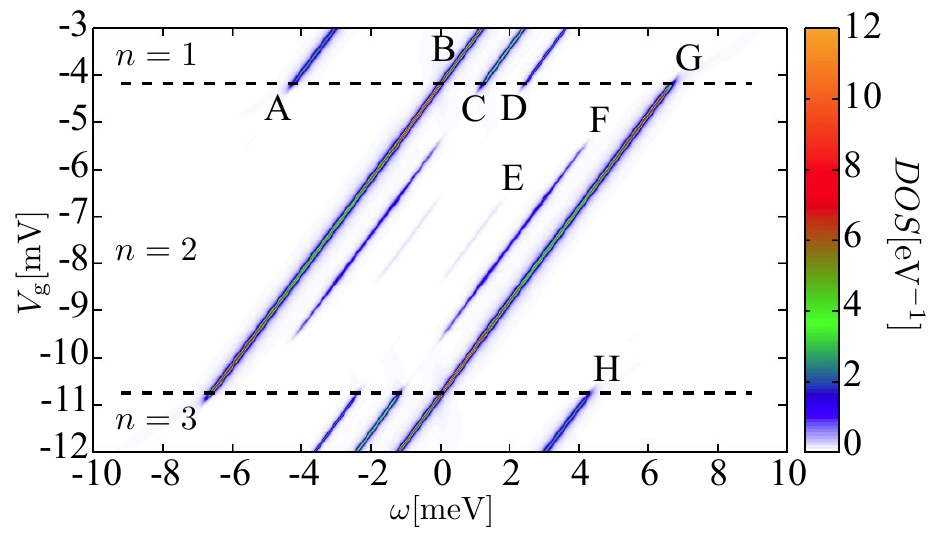}
    \caption{Local spectral densities for $\vb=0$ and for different values of $\vg$ from $-3$~mV to $-12$~mV. The parameters are $U =6$~meV, $J=J^\prime =0.6$~meV, $\Gamma=0.015$~meV and $k_BT=0.06$~meV. Energy $\omega$ is defined with respect to the shared value of the lead chemical potentials.
    \label{FigDOSL}}
\end{center}
\end{figure}
To investigate the importance of both exchange coupling $J$ and pair hopping $J^{\prime}$, we first examine the equilibrium ($\vb=0$) behavior, by calculating  the dot spectral density for various gate voltage $\vg$. Results are shown in Fig.~\ref{FigDOSL}. Varying $\vg$ from $-3$~mV to $-12$~mV, the system shifts from quarter-filled ($n=1$) to three-quarter-filled ($n=3$). 
A remarkable property is the structure of the upper Hubbard band for $n=1$: similarly to the model described in detail by Lee and Phillips~\cite{LP11}, we find for the  quantum dot under study a characteristic splitting of the upper Hubbard band leading to three peaks whose spectral weights decrease in the proportions of $1.5 / 1 / 0.5$ (respectively B, C and D in the figure). The dot remains quarter-filled as long as $\vg$ is higher than $-4.2$~mV. The Coulomb blockade insulating gap is constant and equal to $U-2J-J^\prime$; B and C (C and D) are separated by $2J$ ($2J^\prime$). 
The peak positions and weights of the whole figure can be understood by writing down the  eigenstates of the local Hamiltonian $H_{\rm d}$. Transition energies between these states are reported in Table~\ref{tab} and correspond directly to the peak locations, while their amplitudes are related to the degeneracy. Beyond quarter-filling, peak weights are also affected by double and triple occupancy.
\begin{table}[htb]
\begin{center}
\begin{tabular}{|c|c|c|c|}
\hline
$n_1\rightarrow n_2$ & Transition & Energy$-\varepsilon_0$   \\ 
\hline 
\hline 
A:  $0\rightarrow1$ & $|0,0\rangle \rightarrow |1\rangle$  & 0  \\ 
\hline
\hline
B:  $1\rightarrow2$ &  $ |1\rangle\rightarrow|\sigma,\sigma\rangle$  &  $U-2J-J^\prime$\\ 
 $~~~~1\rightarrow2$ &   $|1\rangle\rightarrow(|\uparrow,\downarrow\rangle+|\downarrow,\uparrow\rangle)/\sqrt{2}$  &  $U-2J-J^\prime$\\ 
\hline 
C:  $1\rightarrow2$ &   $ |1\rangle\rightarrow(|\uparrow,\downarrow\rangle-|\downarrow,\uparrow\rangle)/\sqrt{2}$   &  $U-J^\prime$\\ 
$~~~~1\rightarrow2$ &   $|1\rangle\rightarrow(|\uparrow\downarrow,0\rangle-|0,\uparrow\downarrow\rangle)/\sqrt{2}$  &  $U-J^\prime$\\ 
\hline 
D:  $1\rightarrow2$ &  $|1\rangle\rightarrow(|\uparrow\downarrow,0\rangle+|0,\uparrow\downarrow\rangle)/\sqrt{2}$ &  $U+J^\prime$\\ 
\hline
\hline
E:  $2\rightarrow3$ &  $(|\uparrow\downarrow,0\rangle+|0,\uparrow\downarrow\rangle)/\sqrt{2}\rightarrow|3\rangle$  &  $2U-3J-3J^\prime$\\ 
\hline
F:  $2\rightarrow3$ &  $(|\uparrow\downarrow,0\rangle-|0,\uparrow\downarrow\rangle)/\sqrt{2}\rightarrow|3\rangle$  &  $2U-3J-J^\prime$\\ 
$~~~~2\rightarrow3$ &  $(|\uparrow,\downarrow\rangle-|\downarrow,\uparrow\rangle)/\sqrt{2}\rightarrow|3\rangle$  &  $2U-3J-J^\prime$\\ 
\hline
G:  $2\rightarrow3$ &  $(|\uparrow,\downarrow\rangle+|\downarrow,\uparrow\rangle)/\sqrt{2}\rightarrow|3\rangle$  &  $2U-J-J^\prime$\\ 
$~~~~2\rightarrow3$ &  $|\sigma,\sigma\rangle\rightarrow|3\rangle$  &  $2U-J-J^\prime$\\ 
\hline
\hline
H:  $3\rightarrow4$ &  $|3\rangle\rightarrow|\uparrow\downarrow,\uparrow\downarrow\rangle$  &  $3U-3J-2J^\prime$\\ 
\hline 
\end{tabular} 
\caption{\label{tab}Transition energies between local  eigenstates of  $H_{\rm d}$ for two orbitals. Hund's coupling and pair hopping partially remove the degeneracy of doubly occupied states. $|1\rangle$ represents any state with one electron and $|3\rangle$ represents any 3-electron state. 
Letters from A to H refer to Fig.~\ref{FigDOSL}.}
\end{center}
\vspace{-0.6cm} 
\end{table}

Upon decreasing $\vg$, we observe, together with a sharp increase of $n$,  a significant transfer of spectral weight between the different bands. SWT is a general property of correlated systems~\cite{MES93,LA02}, in contrast to the rigid band patterns of uncorrelated ones, and occurs even for the one orbital case.
In case of orbital degeneracy,  SWT is more sophisticated because of additional orbital degrees of freedom. 
As shown in Fig.~\ref{FigDOSL}, $J$ and $J^\prime$ strongly affect the transfer which arises sequentially when transitions involving two-electron states (B, C, D) cross the Fermi level: the first crossing (B) triggers the simultaneous disappearance of bands A, C and D.
The corresponding spectral weight is redistributed between  B and G, causing a sudden increase of B amplitude.
Further increasing $\vg$, B amplitude is reduced  by the resurgence of C and D. This non-monotonic SWT has no equivalent in the absence of $J$ and $J^{\prime}$.
This  behavior obtained  for $\vb=0$ will undoubtedly have major implications on the transport properties in the finite bias regime. 
Moreover, a finite $\vb$ will lead to additional SWT.
This will be highlighted in the following results for spectral densities, excitation spectrum and stability diagrams. 

\subsection{Nonlinear transport and excitation spectrum}
\begin{figure}
\begin{center}
    \includegraphics[width=8.5cm]{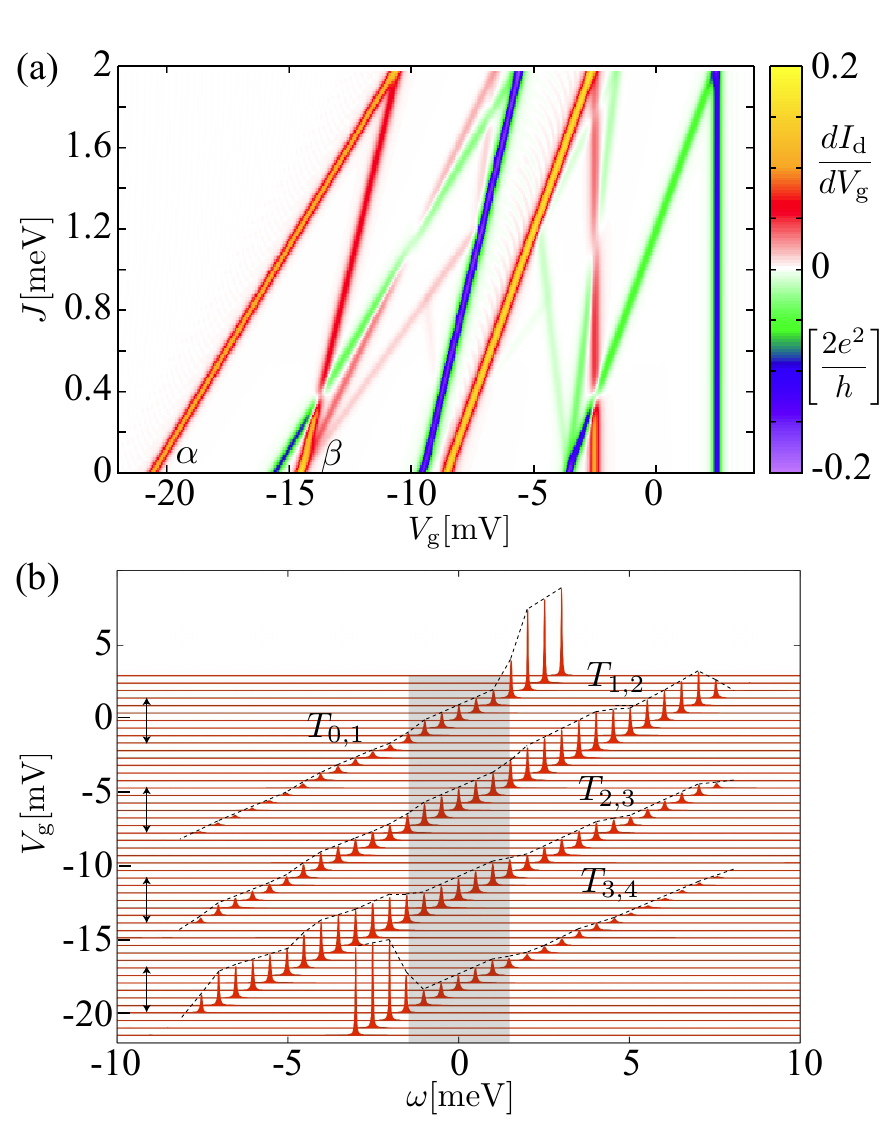}
    \caption{
    (a) Excitation spectrum diagrams $dI_{\rm d}/{d\vg}$ with respect to $J=J^\prime$ and $\vg$. Other parameters are $U =6$~meV, $\Gamma =0.015$~meV, $\vb = 5$~mV and $k_BT=0.06$~meV. (b) Spectral densities for $\vg$ going from $3$~mV to $-21$~mV, $J=J^\prime =0$ and  $\vb = 3$~mV. Energy $\omega$ is defined with respect to the average of $\mu_{L}$ and $\mu_{R}$. $T_{m,n}$ corresponds to the transition between two local states containing respectively $m$ and $n$ electrons. The conducting regime (indicated by double arrows) is achieved when a peak lies inside the bias window.}
\label{figexcitL}\end{center}
\end{figure}
Let us turn to finite bias regime, first calculating 
the excitation spectrum $dI_{\rm d}/{d\vg}$. This quantity is  experimentally accessible~\cite{MFS05} and allows direct visualization of the different transport channels. To reveal the influence of $J$ and $J^\prime$, we plot the excitation spectrum as a function of $\vg$ and $J$. Results obtained for $\vb = 5$~mV are displayed in Fig.~\ref{figexcitL}(a). 
The various drifts upon raising $J$ can be understood by looking at the last column of Table~\ref{tab}. Due to the identity $J^\prime=J$, these drifts range from $-6J$ to $+J$, leading to several crossings.
For increasing $\vg$, positive contributions correspond to the entering of a spectral density peak  into the bias window $[-\vb/2,\vb/2]$, whereas negative ones correspond to peaks leaving this window.  A given peak therefore generates a pair of parallel lines, one positive and the other  negative, separated by  $\vb$. For $J=0$, these pairs are spaced by $U$.
Note that  positive contributions display a great diversity of amplitudes, some of them being nearly vanishing. This can be understood by studying spectral densities of Fig.~\ref{figexcitL}(b), where for the sake of clarity, we  choose $J=J^\prime=0$ in order to have fewer peaks, and $\vb = 3$~mV (far from $U$)   to obtain clearly separated  {\em entering in} and {\em exit from} bias window (light grey in the figure). 
Upon increasing $\vg$, the peak $T_{3,4}$ enters the bias window for  $\vg=-3U-\vb/2$, and leaves it $\vb$ further. A similar process is followed by the peaks $T_{2,3}$, $T_{1,2}$ and $T_{0,1}$, but their amplitudes in the bias window are different, resulting in different values of the current through the dot: this accounts for the magnitude difference between points $\alpha$ and $\beta$ for example, in Fig.~\ref{figexcitL}(a).

Furthermore, Fig.~\ref{figexcitL}(b) clearly shows a significant  SWT upon tuning $\vg$.  
The  spectral density for $\vg =3$~mV corresponds to a number of electron equal to zero. Thus,  the  weight is concentrated in a single peak corresponding to the transition between the empty state and the singly occupied state $T_{0,1}$. 
Unlike the equilibrium case in which  SWT occurs when a structure crosses the Fermi level~\cite{LA02}, a finite bias two-stage SWT  happens  when a peak crosses one of the two chemical potentials $\mu_L$ or $\mu_R$.  Apart from these coincidences, the spectral weights remain constant. This leads to a dot occupation presenting the appearance of a series of plateaus, whose values are not only the integers of the equilibrium case.

\subsection{Stability diagrams}
\begin{figure}
\begin{center}
   \includegraphics[width=8.5cm]{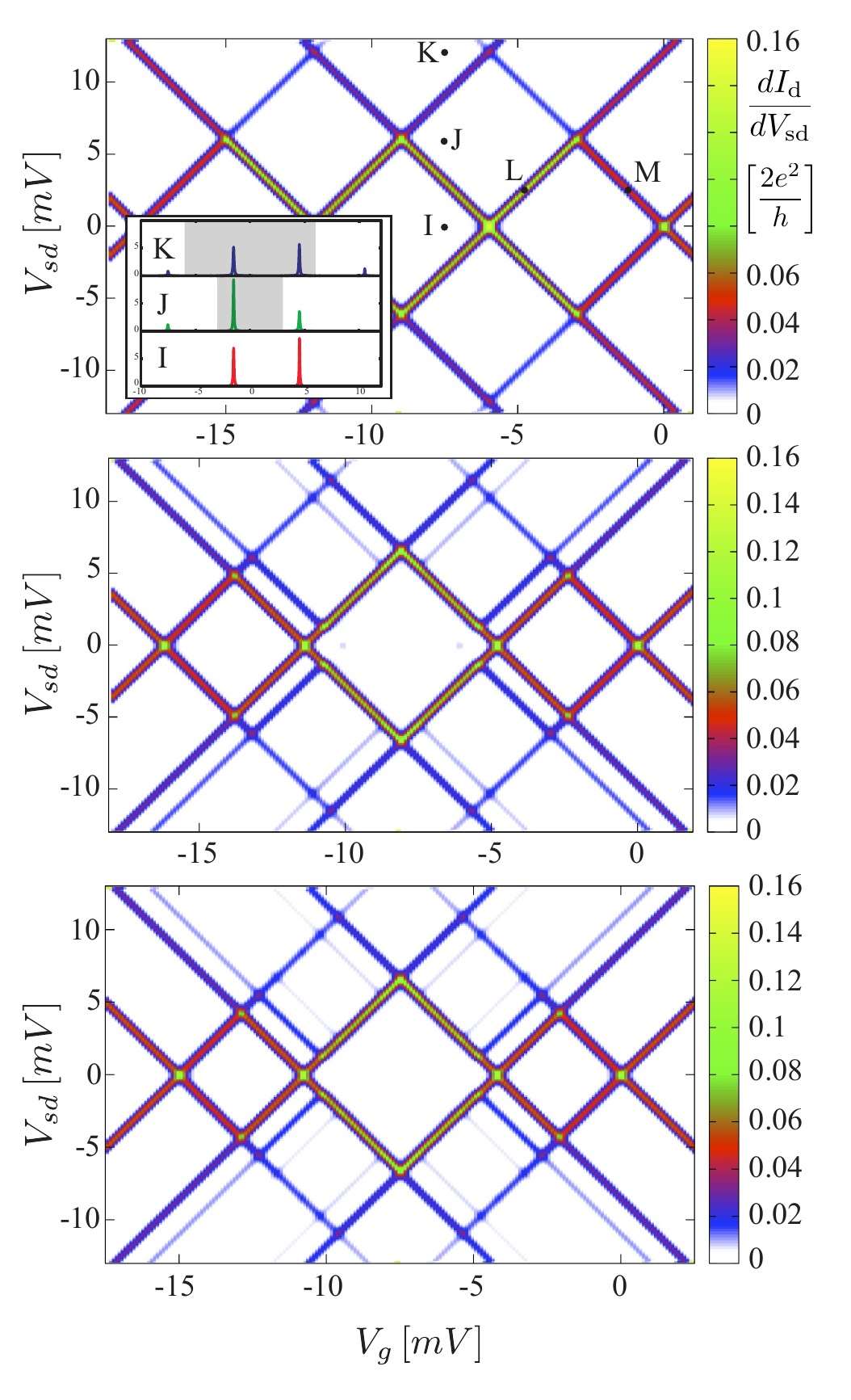}
    \caption{Stability diagrams (a) for $J=J^\prime =0$, (b) for $J= 0.6$~meV and $J^\prime =0$, (c) for $J=J^\prime = 0.6$~meV. Other parameters are $U =6$~meV, $\Gamma =0.015$~meV and $k_BT=0.06$~meV. The inset in (a) displays spectral density corresponding to I, J, K. 
    \label{FigStabL}}
\end{center}
\end{figure}
Stability diagrams are obtained by representing the differential conductance ${dI_{\rm d}}/{d\vb}$ as a function of $\vg$ and $\vb$. They are shown in Fig.~\ref{FigStabL}. In the Coulomb blockade regime, stability diagrams exhibit well-known diamond-shaped structures.
For $J = J '= 0$, in Fig.~\ref{FigStabL}(a), the stability diagrams are composed of eight  lines forming Coulomb diamonds whose diagonal lengths are $2U$ and $U$. The central diamond corresponds to $n = 2$ and the two that surround it to $n = 3$ (left) and $n = 1$ (right). 
The average dot occupancies above and below these diamonds are non-integer. 
Along the diamond lines, the amplitude is highly variable. For example, the amplitude is close to $0.088$ in L and to $0.055$ in M. This is a consequence of the large weight asymmetry between the upper and lower Hubbard bands for $n = 1$. 
The inset with spectral density corresponding to I, J and K, enables to understand how SWT drives Coulomb diamond amplitudes at constant $\vg$: from I to J the structure entering the bias window  (in light grey in the inset) increases significantly, leading to an important current jump; while from J to K the corresponding SWT occurs mainly between two structures inside the window, with a weaker influence on the current variation. 
For $J\neq 0$ and $J'=0$, Fig.~\ref{FigStabL}(b), the central diamond is wider (corresponding to a gap $U+J$) and the side diamonds simultaneously shrink (gap $U-2J$).
For $J'=J\neq 0$, Fig.~\ref{FigStabL}(c), new structures appear and the narrowing of side diamonds is more pronounced ($U-3J$). 
The hierarchy of amplitude between the different lines is directly related to the energy differences between doubly occupied states (see Fig.~\ref{FigDOSL} and Table~\ref{tab}).

\begin{figure}
\begin{center}
   \includegraphics[width=8.5cm]{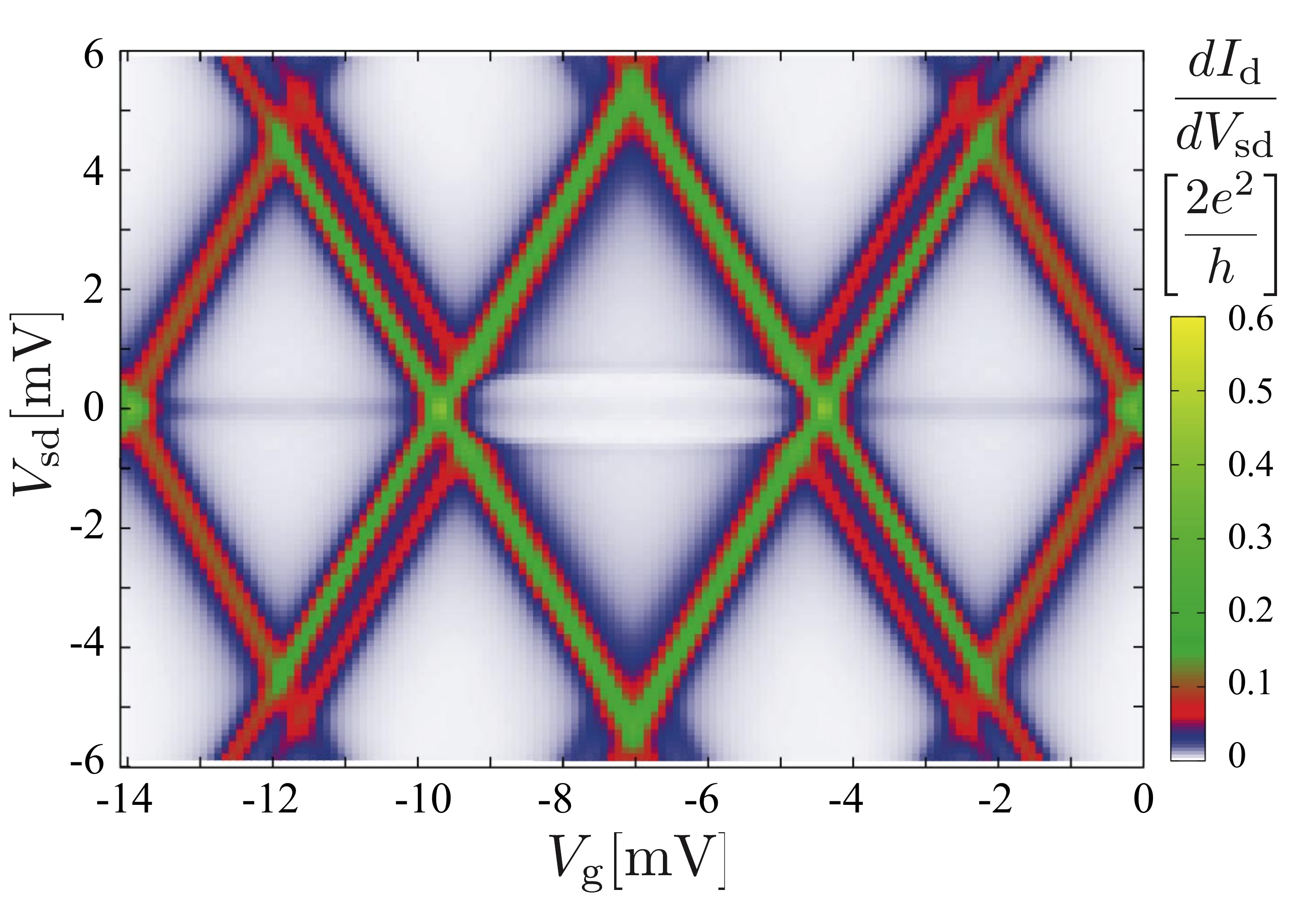}
    \caption{Low temperature stability diagram  for $J= 0.3$~meV, $J^\prime =0$,  $U = 5$~meV, $\Gamma =0.05$~meV, and $k_BT=0.01$~meV. 
        \label{cotunelingL}}
\end{center}
\end{figure}

The central part of stability diagram at lower temperature, revealing the cotunneling regime, is  displayed in Fig.~\ref{cotunelingL} for $J\neq 0$ and $J'=0$. 
The parameters, different from those of Fig.~\ref{FigStabL}(b), have been selected to enhance the visibility of this phenomenon.
In the central diamond ($n=2$), we observe a clear occurrence of inelastic cotunneling processes displaying a characteristic threshold~\cite{DSE2001}. This threshold is given by the energy difference  between two-particle states, namely $2J$. A similar behavior has been observed using a master equation approach in a double-quantum dot in which the cotunneling threshold is the exchange energy~\cite{GL2004}.
A careful examination of the figure shows a zero-biased conductance enhancement
which is due to the Kondo effect. The Kondo temperature is lower for a spin $S=1$ than for $S=1/2$~\cite{SDE2000}  and therefore the enhancement is more pronounced in side diamonds where $n=1$ and $n=3$ respectively.  

\section{Conclusion}
To summarize, we have shown that  Hund's rule exchange coupling and pair hopping  strongly affect transport properties in the Coulomb blockade regime.
If the multi-orbital  Anderson model in the atomic limit can provide the Cou\-lomb diamond positions, their amplitudes however result from spectral weight transfers. Those can only be determined by a proper treatment of correlations in the out of equilibrium regime, especially since the bias voltage variation itself causes transfer. 
The comparative study of the results obtained with the full Hamiltonian on the one hand, and the Hamiltonian containing only the  Hund's coupling on the other hand, clearly indicates the importance of the pair-hopping term, yet rarely considered. In particular, the analysis of experimental stability diagrams without considering pair-hopping narrowing of side diamonds, leads to overestimate the exchange term.


\begin{thebibliography}{100}
\bibitem{hamiltonian} The general Coulomb Hamiltonian can be expressed as follows: 
$H_{\rm C}=1/2 \sum_{i,j,k,l} V_{ijkl}c^\dagger_{i\sigma}c^\dagger_{j\sigma^\prime}
c_{l\sigma^\prime}c_{k\sigma}$. The intra- and inter-orbital interactions are respectively $U=V_{iiii}$
and $U^\prime=V_{ijij}$, while the Hund and pair-hopping parameters are $J=V_{ijji}$ and $J^\prime=V_{iijj}$.
\end{thebibliography}
\end{document}